# Which stylistic features fool ChatGPT research evaluations?

Kayvan Kousha
Statistical Cybermetrics and Research Evaluation Group, Business School, University of Wolverhampton, UK. https://orcid.org/0000-0003-4827-971X; k.kousha@wlv.ac.uk

Mike Thelwall
School of Information, Journalism and Communication, University of Sheffield, UK. https://orcid.org/0000-0001-6065-205X; m.a.thelwall@sheffield.ac.uk

Large Language Models (LLMs) have the potential to be used to support research evaluation and have a moderate capability to estimate the research quality of a journal article from its title and abstract. This paper assesses whether there are language-related factors unrelated to the quality of the research that influence ChatGPT's scores. Using a dataset of 99,277 journal articles submitted to the UK-wide Research Excellence Framework (REF) 2021 assessments, we calculated several readability indicators from abstracts and correlated them with ChatGPT scores and departmental REF scores. From the results, linguistic complexity and length were more strongly associated with ChatGPT research quality scores than with REF expert scores in many subject areas. Although cause-and-effect was not tested, these results suggest that ChatGPT may be more likely than human experts to reward linguistic complexity, with a potential bias towards longer and less readable abstracts in many fields. The apparent preference of LLMs for complex language is an undesirable feature for practical applications of LLMs for research quality evaluation, unless solutions can be found.

## Introduction

The quality of published academic research is repeatedly assessed in the context of appointments, promotions, and national or local departmental research evaluations such as the UK's Research Excellence Framework (REF). These processes consume a lot of time, and the results can be inaccurate, especially when the evaluators lack the time or expertise to properly assess the outputs assigned to them. Historically, citation-based indicators have been used to support or replace expert evaluations, but several different Large Language Models (LLMs) seem to give results that rank more closely with expert scores (Thelwall, 2025c; Thelwall & Yaghi, 2025; Thelwall & Mohammadi, 2026). Nevertheless, the reason why LLM-based quality predictions align moderately well with human scores is not known, especially because they seem to be more effective when LLMs are fed with just article titles and abstracts than when they are fed with full texts (Thelwall, 2025ab). A similar pattern occurs for REF impact case studies, where ChatGPT scores based on titles and summaries correlate more strongly with expert scores than when using full texts (Kousha & Thelwall, 2025). This raises the suspicion that LLMs are "cheating" in the sense of leveraging document attributes unrelated to research quality. Identifying such attributes is therefore necessary to help understand this key LLM property as well as to assess whether authors might take simple steps to manipulate or fool LLM-based research evaluation systems.

Previous analyses of LLMs for research quality evaluation have tended to focus on obtaining the highest alignment with expert scores. Nevertheless, a few studies have tried to identify or test potential sources of bias. A regression analysis of ChatGPT scores for all REF2021 Units of Assessment (UoAs) found ChatGPT to give higher scores to research from some countries, articles in higher impact journals, and articles with longer abstracts. Of these, the last is most concerning since it seems to be unrelated to the quality of the paper. In contrast, some countries invest more in research and journals with higher citation rates may be expected to, on average, publish better research. Average scores also varied substantially between fields, suggesting a topic bias (Thelwall & Kurt, 2025). A comparison of terms occurring disproportionately in high and low scoring tourism titles/abstracts found that ChatGPT tended to score research with complex statistical methods higher and survey research lower (Nunkoo & Thelwall, 2025). For library and information science, LLMs seemed to give higher scores to papers with more technical terminology, theory, statistics, experiments and algorithms, and the opposite for abstraction and research involving libraries, students, or surveys (Zhu et al., 2026; Thelwall, 2025d). It is not clear whether some or all of these reflect underlying quality differences (as judged by human experts), but they are at least suggestive of potential AI bias. From wider research about LLMs for other tasks, it is known that LLMs can be influenced by the credibility or nationality of the source of the information if they are told it (Germani & Spitale, 2025; Vasu et al., 2025) but may not be influenced much by linguistic styles (Kundu & Barbosa, 2024).

Despite the above findings and the research reviewed below, nothing is known about the relationship between textual features of abstracts and LLM scoring. In response, this article addresses the following research questions.
- RQ1: Which textual features of abstracts (e.g., readability, linguistic complexity, and length) associate with LLM scores?
- RQ2: Are these associations stronger for LLM scores than for REF expert scores?
- RQ3: Are there differences in the answers to the above across subjects (UoAs)?

# Background

Although this paper focuses on research quality, there is little relevant research about the relationship between research quality and abstract linguistic characteristics. Thus, this section reviews relevant research with citation data instead.

## *Abstract length and citation impact*

Abstracts summarise the aims, methods, and findings of academic research. Along with the title, they are usually the first part of a paper read to decide whether a study is relevant. Because of this important characteristic, many scientometrics studies have used abstracts as the main source of textual data to investigate associations with research quality or citation impact. Some features of abstracts, such as their length, linguistic complexity, and readability, may influence how frequently papers are cited (Kousha & Thelwall, 2024).

Several studies have found that longer abstracts are associated with higher citation impact, perhaps because longer abstracts contain more information, helping readers to understand key aspects of the research and subsequently making the papers more citable. For example, a study of The Lancet articles (1997–1999) found that highly cited papers had much longer abstracts that contained 2.5 to seven times more words than poorly cited papers (Kostoff, 2007). A large study of one million abstracts across eight science fields found that shorter abstracts (with fewer words and sentences) were associated with fewer citations, although shorter sentences were associated with more citations in mathematics and physics, suggesting disciplinary differences (Weinberger et al., 2015). Another large study of 4.3 million papers from over 1500 journals also found that abstract length positively correlated with citation counts in almost all journals (Sienkiewicz & Altmann, 2016). Other studies across different subjects have similar findings (Didegah & Thelwall, 2013; Hafeez et al., 2019; Sohrabi & Iraj, 2017; Van Wesel et al., 2014). However, experimental evidence suggests that longer abstracts do not necessarily increase reader attention (Helbach et al., 2025). Hence, in the context of research quality assessment, it is not clear whether experts or other evaluators would be influenced by abstract length when judging research quality.

### Abstract readability and citation impact

Many studies have investigated whether more readable abstracts (e.g., Flesch reading ease score) associate with higher citation counts, finding that more complex (less readable) abstracts receive more citations, albeit with some disciplinary differences. For example, abstracts of highly cited information science articles are less readable (as measured by the Flesch Reading Ease score) (Gazni, 2011). A very large study of 4.3 million journal articles found that abstract complexity (measured by the Gunning Fog Index) positively associated with citation counts (Sienkiewicz & Altmann, 2016). Another study found that abstract readability, measured by the Flesch Reading Ease score, negatively correlated with citation counts in Biology and Biochemistry, while no significant association was found in Chemistry or Social Sciences (Didegah & Thelwall, 2013). Similarly, a study comparing highly cited and uncited Web of Science papers across multiple scientific fields found that abstracts of highly cited articles were longer, more linguistically complex, and less readable (Hu, Wang, & Deng, 2021). For articles on emerging technologies, less readable abstracts, measured using several readability indicators (the Flesch Reading Ease, Flesch–Kincaid Grade Level, and Gunning Fog Index), associated with higher citation impact (Ante, 2022). A study of 71,628 abstracts in Language and Linguistics journals (1991–2020) also found that lower abstract readability associated with higher citation counts, although the relationship was weak (Wang, Liu, & Zhou, 2022). A smaller-scale study of 550 articles published in Science also reported that abstract readability broadly related to online attention (Altmetric scores) received by articles (Jin et al., 2021).

## Methods

The research design was to obtain LLM research quality scores and stylistic characteristics of abstracts, such as readability, linguistic complexity, and length, for a large set of academic journal articles across different subjects and identify any relationship between the two. The second part of the research design was to repeat the process for expert scores instead of LLM scores, to assess whether there were any differences between experts and LLMs in the patterns found. Such differences would be suggestive of AI bias rather than indirect factors associating with research quality. The third part of the research design (not addressing a research question but added for context) involved replacing LLM scores with citation impact indicators to assess whether stylistic characteristics of abstracts were similarly or differently related to citation impact.

### Dataset

The only recent large-scale dataset of published journal articles from all fields with an associated quality indicator is the UK REF2021 output set. This is a collection of research outputs that were submitted for evaluation. Each output had been assigned a research quality score on an integer scale of 1 to 4 by at least two field experts from the appropriate UoA broadly field-based grouping (https://2021.ref.ac.uk/publications-and-reports/panel-criteria-and-working-methods-201902/). Whilst the individual output scores have been destroyed, the number of outputs awarded each of the four score levels for each "submission" is public. A "submission" can be thought of as approximating a department or school in a single university, although it does not have to be. The term "department" will be used here for convenience. Each output was assigned the average score from the department that submitted it, as a proxy quality rating (or the mean of all submitting departments, if an article was submitted by multiple authors). Since every department in REF2021 had multiple different scores, this is inaccurate in all cases. Nevertheless, since individual output scores correlate with departmental scores, in the absence of bias, this serves to dampen correlations but not to invalidate any positive correlations.

Individual scores have been generated for 500 articles of each of UoAs 1 to 6 for a previous paper (Thelwall & Mohammadi, 2026) and these were used here. They were obtained by the second author of the current article and represent his estimate of the REF quality score of these articles.

A spreadsheet of outputs assessed was obtained from the REF2021 website (https://results2021.ref.ac.uk/all-submission-data) and cross-referenced with a spreadsheet of departmental scores (https://results2021.ref.ac.uk/profiles/export-all), via the institution and submission (department) columns. An undesirable feature of the scores is that they are fully public and may have been included in the LLM training data. Nevertheless, it seems unlikely that they were because the data is in Excel format and therefore not a natural LLM input. In addition, cross-referencing the two spreadsheets is needed to associate an article with a score, which also seems unlikely. Finally, previous tests have shown that the API version of ChatGPT rarely knows the department or even journal of an article from its title and abstract (Thelwall, 2026), so is unlikely to be able to lookup REF-related quality information about articles.

For this dataset, in each UoA, non-article outputs (e.g., books, chapters) and the articles with the shortest 10% of abstracts were discarded. For the latter, many short abstracts were from short form articles (e.g., letters, correspondence) or were incomplete. There were too many articles to check individually, especially for the difficult decision of deciding whether an article was a short form contribution, and so the 10% heuristic was used as a blunt instrument to remove unsuitable articles. The final dataset size was 99,277 unique journal articles without short abstracts (106,843 including duplicates).

### Citation rates

A field and year normalised citation rate was calculated for each article as an additional indicator. The Normalised Log-transformed Citation Score (NLCS) was used, which adjusts for the field and year of publication as well as for the highly skewed nature of citation data (Thelwall, 2017). For this calculation, all citation counts were first transformed with $\ln(1+x)$ to reduce skewing, and then each article's (log transformed) citation count was divided by the average log-transformed citation count of all articles from the same field and year (averaging the values from multiple fields when relevant). The data for this were taken from Scopus in 2024, giving at least 3 years of citations for each paper. The fields used for the normalisation were the All Science Journal Classification (ASJC) scheme used by Scopus. This process reduces age and field biases in citation data.

## LLM scores

Scores were obtained from multiple LLMs, all recycled from a previous paper (Thelwall, 2025c). The details are in that paper but, in brief, the models were ChatGPT-4o, its smaller sibling ChatGPT-4o mini, and ChatGPT-5 mini. For each model, the articles were submitted to ChatGPT through its Applications Programming Interface (API).

The scores were based on a system prompt describing the task with the instructions originally given to the human reviewers in REF2021, requesting a score for the article based on its title and abstract. The scores follow the REF2021 scale of 1* (nationally relevant) to 4* (world leading) (for exact prompts, see: Thelwall, 2025c). In all cases except one (described next), the scores were the average of five separate requests because averaging has been shown to increase the correlation with expert scores (Thelwall, 2025abc).

The final case without five repetitions used a simple structured prompt requesting separate comments on the three quality components of originality, rigour, and significance, followed by a score for each one, and then an overall comment and an overall score at the end (for exact prompts, see: Thelwall, 2025c). This is similar to the process that experts might be requested to use, except that the user prompt states that fractional scores (e.g., 3.5*) are allowed, whereas the human experts had to assign integers: 1*, 2*, 3* or 4*.

For the primary analysis, all scores were averaged to give a single number representing the combined estimate from both ChatGPT-4o, ChatGPT-4o mini and ChatGPT-5 mini. The previous study had found this combination to give the best overall quality ranking, in terms of the strength of correlation with expert scores (Thelwall, 2025c). The increased power of averaging older and newer models might derive from the additional iterations or the combination of perspectives from the different models.

## Readability features

Abstract readability and textual characteristics of abstracts in the dataset were calculated with a Python script based on the *textstat* library (https://pypi.org/project/textstat/). For each abstract, several standard readability indicators were extracted: Flesch Reading Ease, Flesch–Kincaid Grade Level, Gunning Fog, SMOG, and ARI. Abstract length was also calculated since this has previously been shown to associate with LLM scores (Thelwall & Kurt, 2025). The main analysis focuses on the Flesch Reading Ease and Flesch–Kincaid Grade Level measures because they are widely used readability indicators in scientometrics research. The Flesch Reading Ease score estimates how easy a text is to read based on average sentence length and syllables per word, with higher scores indicating easier texts (Flesch, 1948). The Flesch–Kincaid Grade Level uses the same two linguistic components but expresses the result as a U.S. school grade level required to understand the text (Kincaid et al., 1975). The main analysis focuses on the Flesch–Kincaid Grade Level because it is widely used. Additional readability indicators (Flesch Reading Ease and Gunning Fog) are reported in the appendix (Figures S1 and S2, respectively).

## Textual features

An enormous range of textual indicators can be calculated. For this study, only indicators investigated in previous citation-analysis studies were used to keep the calculations manageable and to avoid adding many indicators speculatively. Testing too many indicators without a theoretical justification risks finding spurious positives through testing volume (the Bonferroni familywise error rate issue). On this basis, the following indicators were chosen for the main analyses: abstract length (word count), words per sentence, and syllables per word. These represent key aspects of abstract style, as used in standard readability indicators, and length. To avoid redundancy, the many closely related indicators (e.g., characters per word, characters per sentence, or total number of characters) are not reported.

### *Analysis*

Previous studies with LLM scores of academic outputs have tended to find that the individual scores are usually wrong, with a high preference for 3*, but that the useful information is the ranking of the scores, which tends to align moderately well with expert scores (slightly weaker with a quality proxy) (Thelwall, 2025c). Thus, the only relevant measure of usefulness is rank correlation. Whilst there are several different rank correlations, Spearman was chosen in preference to Kendall's tau because the latter does not penalise error size. To report statistical significance, bootstrapped 95% confidence intervals were calculated for all Spearman correlations. These are not shown in the graphs because they would be difficult to read for with error bars.

## Results

### *Main Panel A: Health and life sciences (article-level comparison)*

For the health and life sciences (UoAs 1 to 6) for which there are 500 individual scores per UoA (n=3,000, including duplicate articles that are in multiple UoAs), it is possible to directly compare the ChatGPT scores (rather than departmental averages) and individual article scores with readability indicators. Overall, the most consistent Spearman correlations are for syllables per word and words per sentence and the Flesch-Kincaid Grade Level combination of the two. In all cases the ChatGPT correlations are more positive than the REF expert correlations, suggesting that ChatGPT may be more susceptible to these factors, although causation is unproven. For example, in Clinical Medicine the Spearman correlation between Flesch–Kincaid Grade Level and ChatGPT scores was 0.329 compared with 0.124 for REF expert scores. Similar differences occur for Public Health (0.218 vs 0.037), Allied Health Professions (0.291 vs 0.072), and Psychology (0.294 vs 0.137). ChatGPT scores were also more strongly correlated with the underlying linguistic features that form this measure, such as words per sentence (0.306 vs 0.150 in Clinical Medicine) and syllables per word (0.220 vs 0.039). Similar patterns occurred for other indicators of lexical complexity, including letters per word (0.234 vs 0.023).

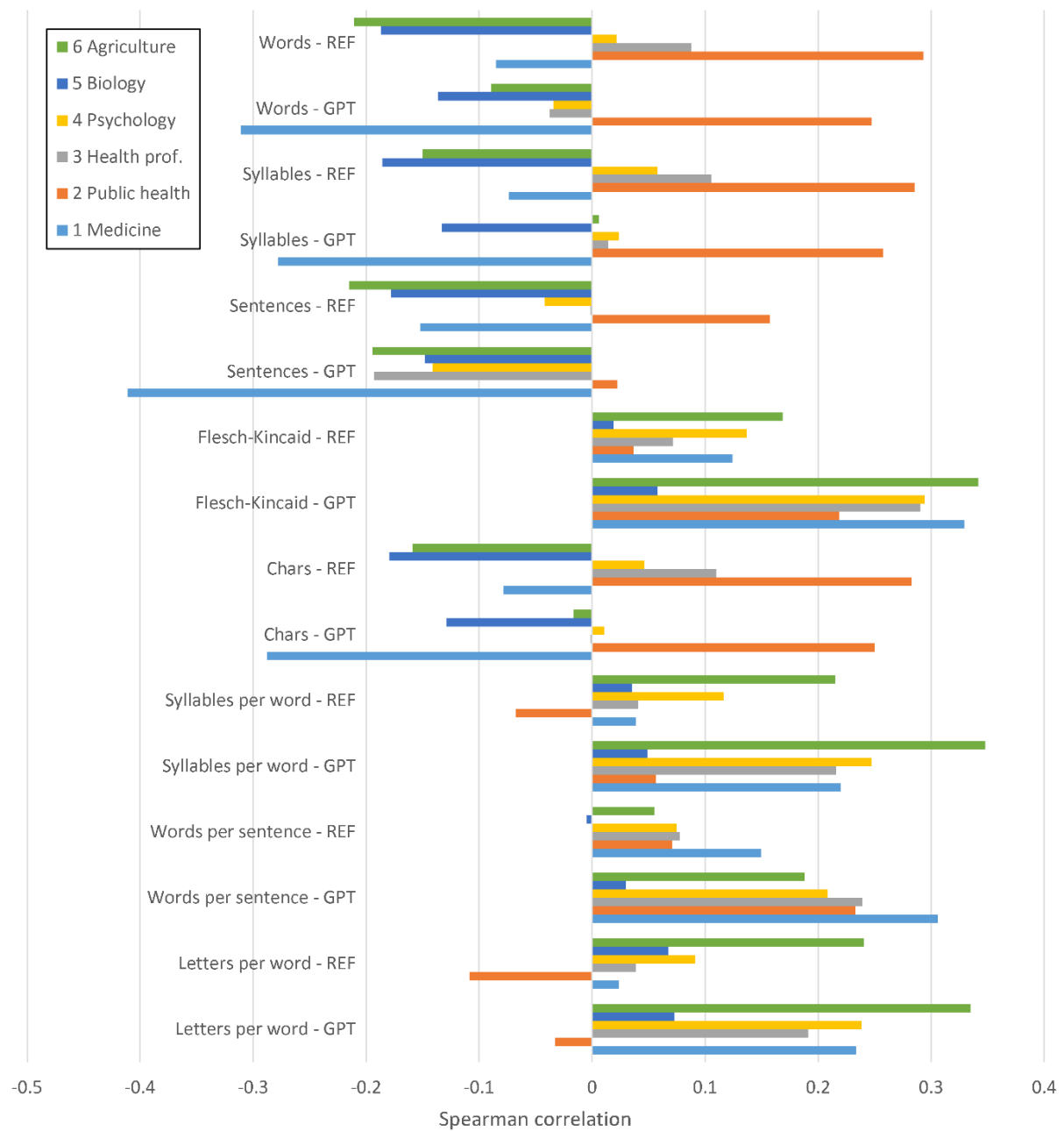

Figure 1. ChatGPT or expert REF scores correlated against text complexity indicators for REF2021 health and life sciences UoAs (n=500 for each UoA).

*Smaller vs. Larger dataset analysis for UoAs 1 to 6*

Because individual REF scores were not available for the larger dataset within Main Panel A (UoAs 1 to 6; n=39,832 articles), departmental average REF scores were used as a research quality proxy for individual papers. To assess whether the patterns in the 500-article per unit analysis (with individual scores) replicate with the proxy indicator, the analysis was repeated using the larger dataset for the six health and life science UoAs with the departmental average REF scores. Tables 1 to 3 compare the correlations between abstract linguistic complexity indicators and evaluation scores for the 500-paper dataset with individual scores and the larger dataset with departmental averages. Across all six health and life science UoAs, the correlations with ChatGPT scores are consistently higher than those for REF scores in both datasets. This pattern remains broadly similar even when REF scores are represented by departmental average proxies in the larger dataset. This suggests that analyses based on departmental average REF scores can be used to make inferences equivalently to analyses based on individual scores.

Table 1. Correlations between the Flesch-Kincaid Grade Level and research quality scores in the 500-paper sample and the larger dataset.

| UoA | ChatGPT (500 papers) | REF individual scores | N | ChatGPT (Larger dataset) | REF departmental score | N |
|---|---|---|---|---|---|---|
| Clinical Medicine | 0.220 | 0.039 | 500 | 0.273 | 0.062 | 9,602 |
| Public Health | 0.056 | -0.067 | 500 | 0.045 | 0.009 | 3,821 |
| Allied Health Professions | 0.216 | 0.041 | 500 | 0.201 | 0.055 | 9,337 |
| Psychology | 0.247 | 0.116 | 500 | 0.195 | 0.085 | 7,881 |
| Biological Sciences | 0.049 | 0.035 | 500 | 0.135 | 0.052 | 6,123 |
| Agriculture | 0.348 | 0.215 | 500 | 0.326 | 0.208 | 3,068 |

Table 2. Correlations between the words per sentence and research quality scores in the 500-paper sample and the larger dataset.

| UoA | ChatGPT (500 papers) | REF individual scores | N | ChatGPT (Larger dataset) | REF departmental score | N |
|---|---|---|---|---|---|---|
| Clinical Medicine | 0.306 | 0.150 | 500 | 0.266 | 0.039 | 9,602 |
| Public Health | 0.233 | 0.071 | 500 | 0.191 | -0.013 | 3,821 |
| Allied Health Professions | 0.239 | 0.077 | 500 | 0.225 | 0.049 | 9,337 |
| Psychology | 0.208 | 0.075 | 500 | 0.160 | 0.000 | 7,881 |
| Biological Sciences | 0.030 | -0.005 | 500 | 0.021 | -0.032 | 6,123 |
| Agriculture | 0.188 | 0.055 | 500 | 0.174 | 0.088 | 3,068 |

Table 3. Correlations between syllables per word and evaluation scores in the 500-paper sample and the larger dataset.

| UoA | ChatGPT (500 papers) | REF individual scores | N | ChatGPT (Larger dataset) | REF departmental score | N |
|---|---|---|---|---|---|---|
| Clinical Medicine | 0.329 | 0.124 | 500 | 0.353 | 0.069 | 9,602 |

| Public Health | 0.218 | 0.037 | 500 | 0.185 | 0.003 | 3,821 |
| Allied Health Professions | 0.291 | 0.072 | 500 | 0.285 | 0.071 | 9,337 |
| Psychology | 0.294 | 0.137 | 500 | 0.244 | 0.057 | 7,881 |
| Biological Sciences | 0.058 | 0.019 | 500 | 0.113 | 0.013 | 6,123 |
| Agriculture | 0.342 | 0.168 | 500 | 0.322 | 0.177 | 3,068 |

## *Analysis across all 34 REF UoAs*

Based on the comparison between ChatGPT and REF expert correlations in Main Panel A (Figure 1), three indicators were selected for further analysis across all 34 REF UoAs: Flesch–Kincaid Grade Level, words per sentence, and word count. These indicators were chosen because they capture different aspects of abstract writing style. Flesch–Kincaid Grade Level represents overall linguistic complexity, words per sentence reflects syntactic complexity, and word count measures overall abstract length, which is widely used in scientometric research. Additional analyses using stylistic indicators with different linguistic characteristics, including Gunning Fog, SMOG, average letters per word, and sentence count, syllabus count are reported in the Appendix.

## *Flesch–Kincaid Grade Level*

The Flesch–Kincaid Grade Level is a weighted combination of syllables per word and words per sentence that estimates the educational grade level required to understand a text. Higher scores indicate more linguistically complex abstracts and therefore a higher level of reading difficulty (Kincaid et al., 1975). Overall, 23 of the 34 UoAs had statistically significant positive correlations between ChatGPT scores and Flesch–Kincaid Grade Level, while one unit had a statistically significant negative association and the remaining UoAs had no statistically significant relationships, possibly due to small sample sizes and/or weak relationships. This indicates that linguistically more complex abstracts tended to receive higher ChatGPT scores. However, REF expert departmental average scores statistically significantly positively correlated with grade level in only 9 UoAs out of 34, following the same direction as ChatGPT, although these correlations were generally much smaller (Figure 2). Six UoAs also had statistically significant negative correlations between grade level and REF expert scores, including Politics and International Studies, Education, and several humanities subjects.

In several biological and physical sciences, ChatGPT scores had stronger positive associations than REF expert scores with abstract complexity features. For example, in Engineering (ChatGPT: 0.215; REF: 0.073), Chemistry (0.212; 0.083), and Biological Sciences (0.113; 0.013). These results suggest that ChatGPT scores were more strongly associated with linguistic complexity in abstracts, whereas REF expert scores were generally weakly related or unrelated to this readability measure.

Across many social science subjects, ChatGPT scores were more likely to increase with higher grade levels, whereas REF expert scores were often weakly or not significantly related to this measure. In several humanities subjects, REF expert scores also had weak and inconsistent associations with Flesch–Kincaid Grade Level, with some significant negative correlations in Modern Languages and Linguistics (–0.163), English Language and Literature

(−0.109), and Philosophy (−0.112), suggesting that more linguistically complex abstracts were not favoured by expert evaluations in these subjects.

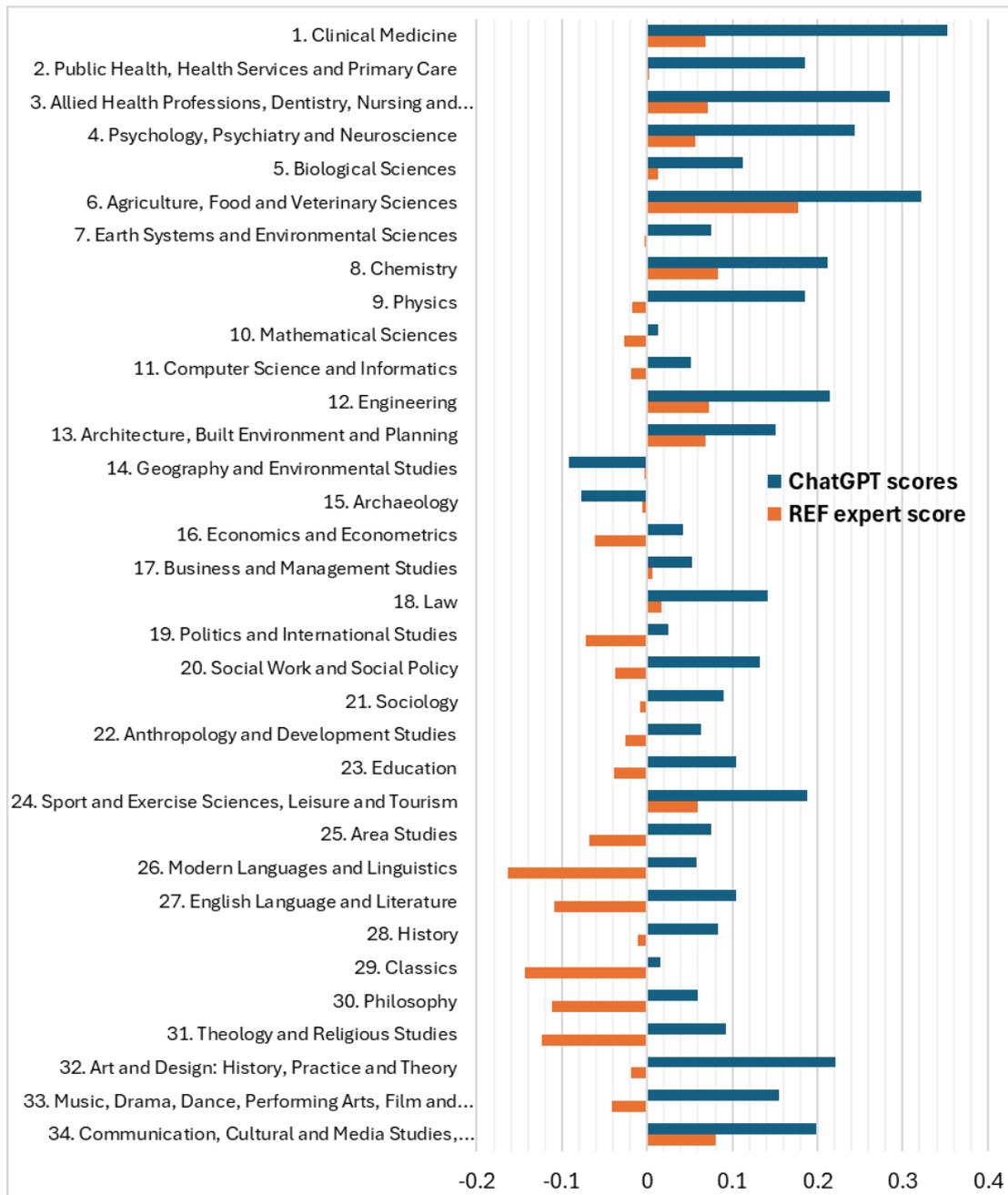

Figure 2. Spearman correlations between Flesch–Kincaid Grade Level and ChatGPT and REF departmental average expert scores across the 34 REF Units of Assessment.

*Abstract length*

In Figure 3, 29 of the 34 UoAs had significant correlations between ChatGPT scores and abstract length (number of words), including 22 positive and 7 negative associations. Positive correlations indicate that longer abstracts tended to receive higher ChatGPT scores, whereas negative correlations suggest the opposite. Negative correlations occurred in several biomedical and STEM fields, such as Clinical Medicine (−0.311) and Biological Sciences (−0.188), whereas positive associations appeared in many social science and humanities subjects, including History (0.218), English Language and Literature (0.194), and Philosophy (0.214). The negative association between abstract length and ChatGPT scores in some biomedical and STEM subjects, especially Clinical Medicine, may be related to the style of abstracts in these fields. Medical journals often require structured abstracts (e.g., Aims, Participants, Methods, Results) and have strict word limits to summarise key findings clearly. However, shorter abstracts do not necessarily mean they are easier to read. As shown in Figure 2, Clinical Medicine also had the strongest positive association between linguistic complexity and ChatGPT scores. In other words, ChatGPT may be more likely to assign higher scores to shorter but more linguistically complex abstracts.

In contrast, departmental average REF expert scores had statistically significant associations with abstract length in only 15 UoAs and most of these were negative (12 negative and 3 positive), indicating that shorter abstracts were often associated with higher average expert scores in these subjects.

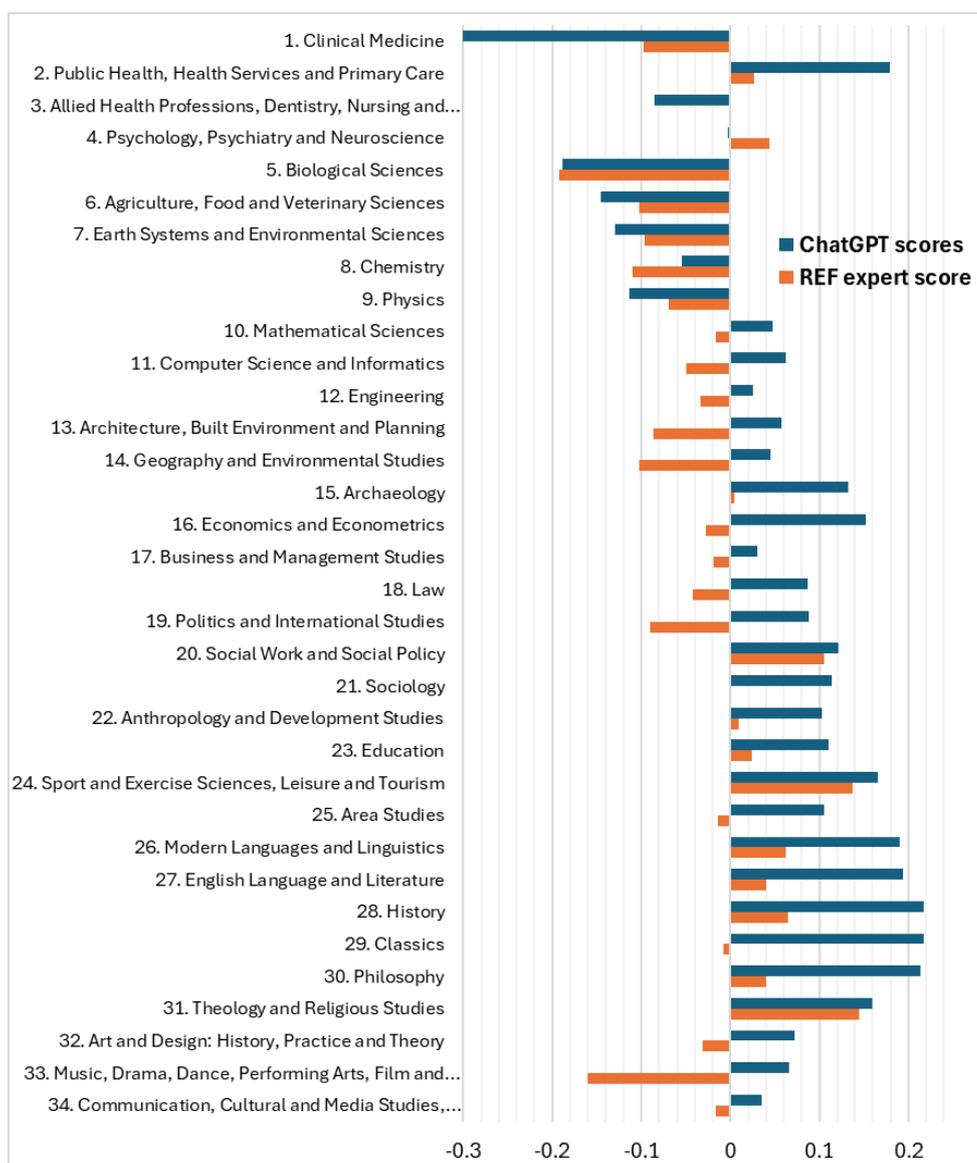

Figure 3. Spearman correlations between abstract length (word count) and ChatGPT and REF departmental average expert scores across the 34 REF Units of Assessment.

*Words per sentence*

Words per sentence partially reflects the syntactic complexity of the text in terms of average sentence length. Many medical and Science, Technology, Engineering and Maths (STEM) UoAs had statistically significant positive correlations between ChatGPT scores and words per sentence, and these were generally stronger and more frequent than those for REF departmental average expert scores (Figure 4). However, in several arts and humanities subjects, REF expert scores had negative associations with sentence length while ChatGPT scores had little or no relationship, for example in Modern Languages and Linguistics (−0.198) and English Language and Literature (−0.135), suggesting that some REF expert evaluations may favour clearer writing with shorter sentences.

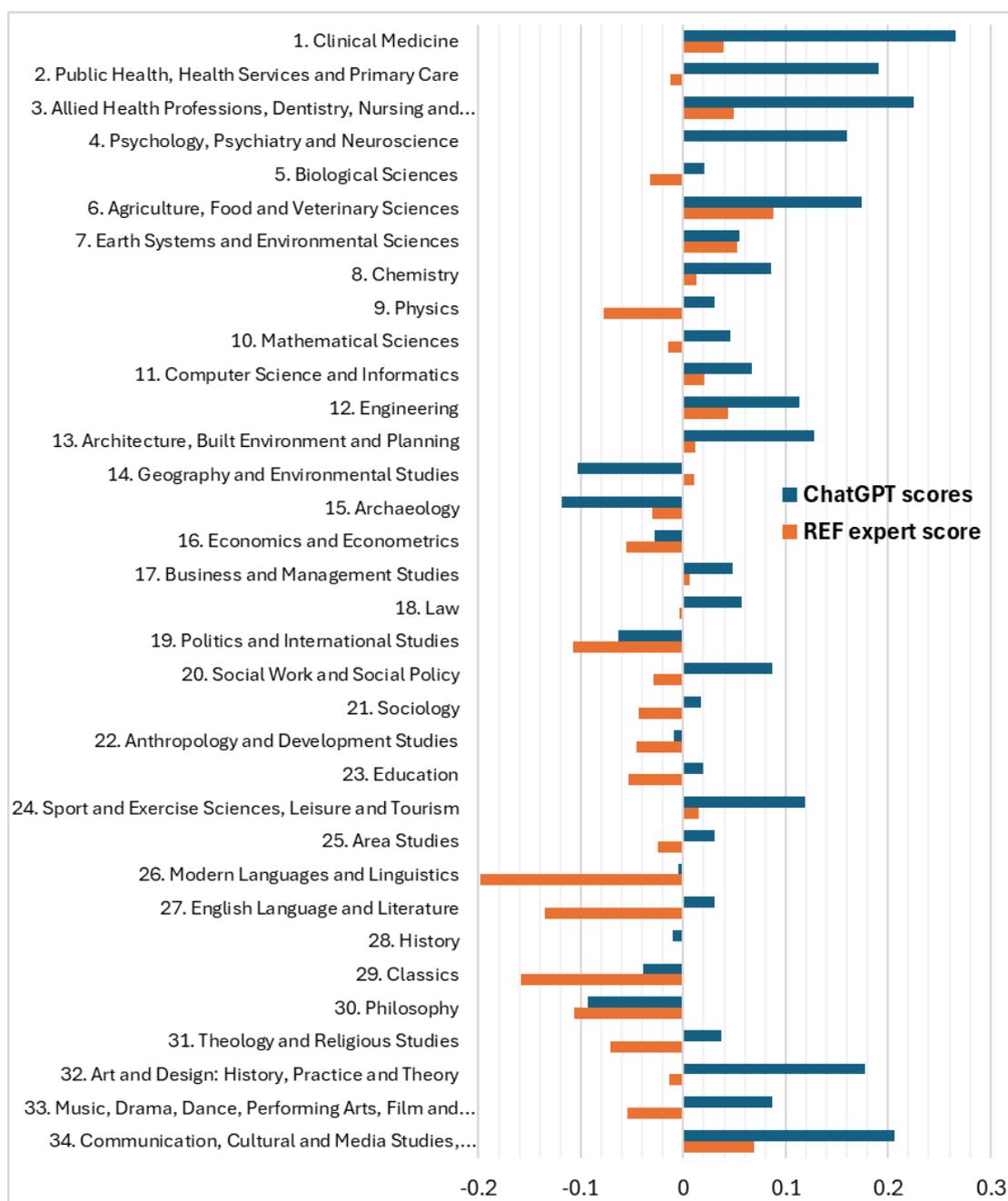

Figure 4. Spearman correlations between words per sentence and ChatGPT and REF expert scores across the 34 REF Units of Assessment.

## Syllables per word

Syllables per word is a simple vocabulary complexity indicator. Higher values generally indicate longer or more technical words in the text. In most subjects, abstracts with more syllables per word tended to receive higher ChatGPT scores compared with REF expert departmental average scores. Across the 34 REF UoAs, 28 had statistically significant positive correlations between ChatGPT scores and syllables per word, whereas expert scores had statistically significant correlations in only 11 UoAs (Figure 5). For example, positive correlations between syllables per word and ChatGPT scores occurred in Clinical Medicine (0.273), Physics (0.262), Chemistry (0.236), and Engineering (0.210), whereas REF expert correlations with syllables per word in these subjects were much weaker. Similar patterns also occurred in several social sciences and humanities subjects, including Philosophy (0.211), History (0.190), Education (0.148), and English Language and Literature (0.145), where REF correlations were generally weak, close to zero, or negative. This suggests that ChatGPT scores may be more sensitive than expert evaluations to vocabulary complexity in abstracts.

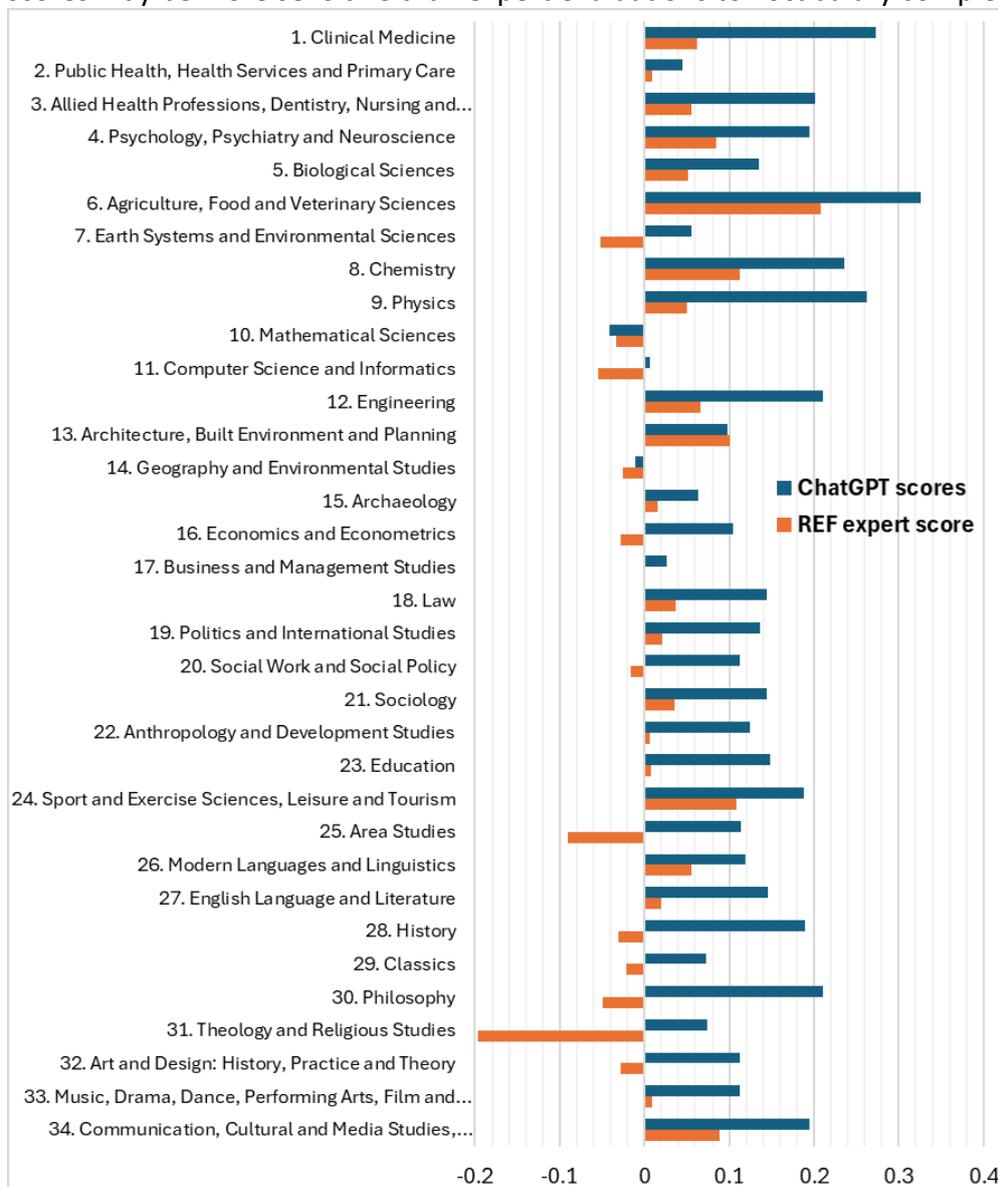

Figure 5. Spearman correlations between syllables per word and ChatGPT and REF expert scores across the 34 REF Units of Assessment.

# Discussion

*Comparison between ChatGPT scores and citation impact*

As a further analysis, ChatGPT scores and the Normalised Log-transformed Citation Score (NLCS 2024) were correlated with abstract readability (Flesch–Kincaid Grade Level). Both ChatGPT scores and NLCS were collected at the individual article level, allowing a direct comparison between the two indicators.

Figure 6 compares the correlations between ChatGPT and citation impact scores with Flesch–Kincaid Grade Level across the 34 REF Units of Assessment. In several biomedical and STEM subjects, ChatGPT scores were clearly associated with higher linguistic complexity, but the same was not true for citation impact: correlations were generally weaker and usually negative. This suggests that LLM-based evaluations are usually more sensitive to abstract complexity than citation-based indicators of research impact.

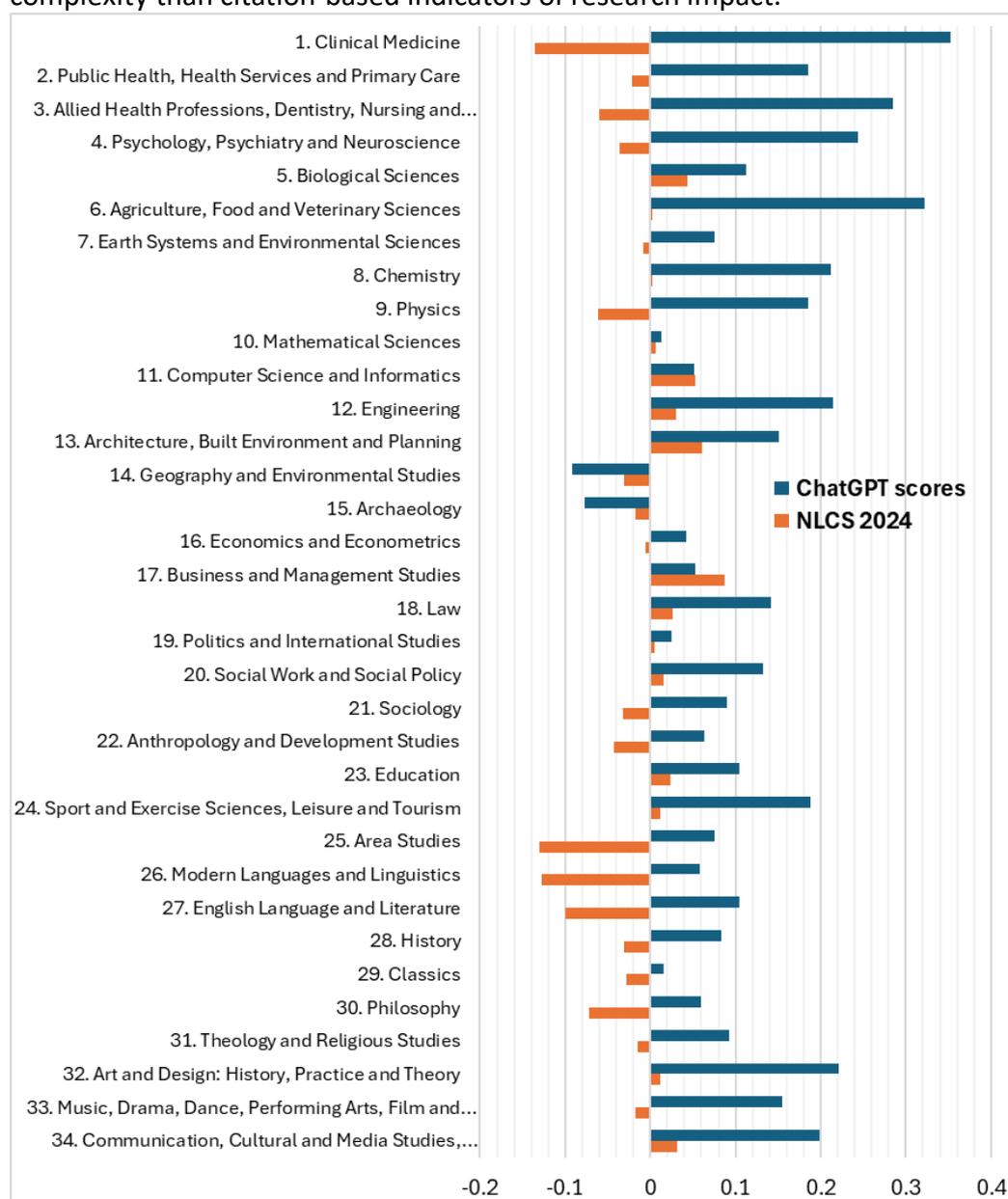

Figure 6. Spearman correlations between ChatGPT scores and citation impact (NLCS 2024) with Flesch–Kincaid Grade Level across the 34 REF Units of Assessment.

*Limitations*

This study has several data, methods, and interpretation limitations. The journal articles are from a single country and there may be different patterns for other countries, and particularly for those where English (the dominant language of science) is not commonly spoken. The articles are also preselected for quality because each department tries to submit its best work for the evaluation. This is likely to have reduced the strength of the correlations. Another limitation is that other LLMs, prompting strategies, or textual indicators may have yielded different results.

Unfortunately, individual REF expert scores were not available for the larger dataset used in Figures 2–4. Instead, departmental average REF scores were used as a proxy for individual papers. This approach reduces variation in the expert scores assigned to articles because multiple papers from the same department receive the same average score. As a result, correlations between abstract characteristics and REF scores may be weaker than if individual expert scores were available, although the direct comparison for UoAs 1 to 6 suggests that the reduction would not be large. Nevertheless, comparisons between ChatGPT and REF correlations should be interpreted cautiously, as ChatGPT scores were available at the article level whereas REF scores represent departmental averages.

Another limitation is that the analysis focuses on key textual and readability features, such as abstract length and linguistic complexity, and does not consider other potentially relevant characteristics such as abstract structure (e.g., background, aims, methods, and conclusions), or methodological detail. These additional textual features may also influence LLM scoring.

Finally, the study used correlation analyses which do not show causal relationships. For example, longer or more linguistically complex abstracts may not directly cause higher ChatGPT scores but may instead be associated with other characteristics of academic writing that influence the model's evaluation.

# Conclusion

Although cause-and-effect has not been proven, the results suggest that abstracts with complex styles impress ChatGPT more than they impress expert evaluators, in the sense of ChatGPT tending to assign them higher scores. This is a plausible finding in the context of the text pattern matching of LLMs. For example, it may have previously ingested many unpublished badly written and poor manuscripts by hopeful authors, therefore "learning" to associate plain grammar with weak research. It is also possible that its historical data, with deliberately complex language being highly regarded, has given the same effect. More generally, it may have identified that outside academia humans can be impressed by complex language.

The findings also raise the possibility that some textual characteristics of abstracts could be intentionally modified to obtain higher LLM evaluation scores. If LLM systems tend to give higher scores to features such as longer or more linguistically complex abstracts, these characteristics could potentially be adjusted by authors to increase their scores. If LLMs were used to support research evaluation processes such as peer review or national assessments like the REF, this could create the motivation to modify writing style rather than improve the underlying research quality. The behavioural consequences of such motivations are not clear yet. Hence, identifying which textual features influence LLM evaluation is important to understand potential risks of manipulation and to support the responsible use of LLMs in research assessment.

Given the results, it seems desirable to reduce the apparent impact of language complexity on LLM scores so that (a) authors are not rewarded for their language instead of their research and (b) authors do not try to write in a more complex style to impress LLMs. This might be achieved by incorporating prompt instructions to ignore language style or by using LLMs trained exclusively on relevant academic texts, if possible. Whilst in theory, the problem could be eradicated by assigning a score penalty to complex texts to reduce their LLM score rank, this may be undesirable interference in the evaluation process and perhaps difficult to get agreement about.

# Acknowledgement

This study is partly funded by the Economic and Social Research Council (ESRC), UK (UKRI2101).

# Appendix: Additional readability analysis

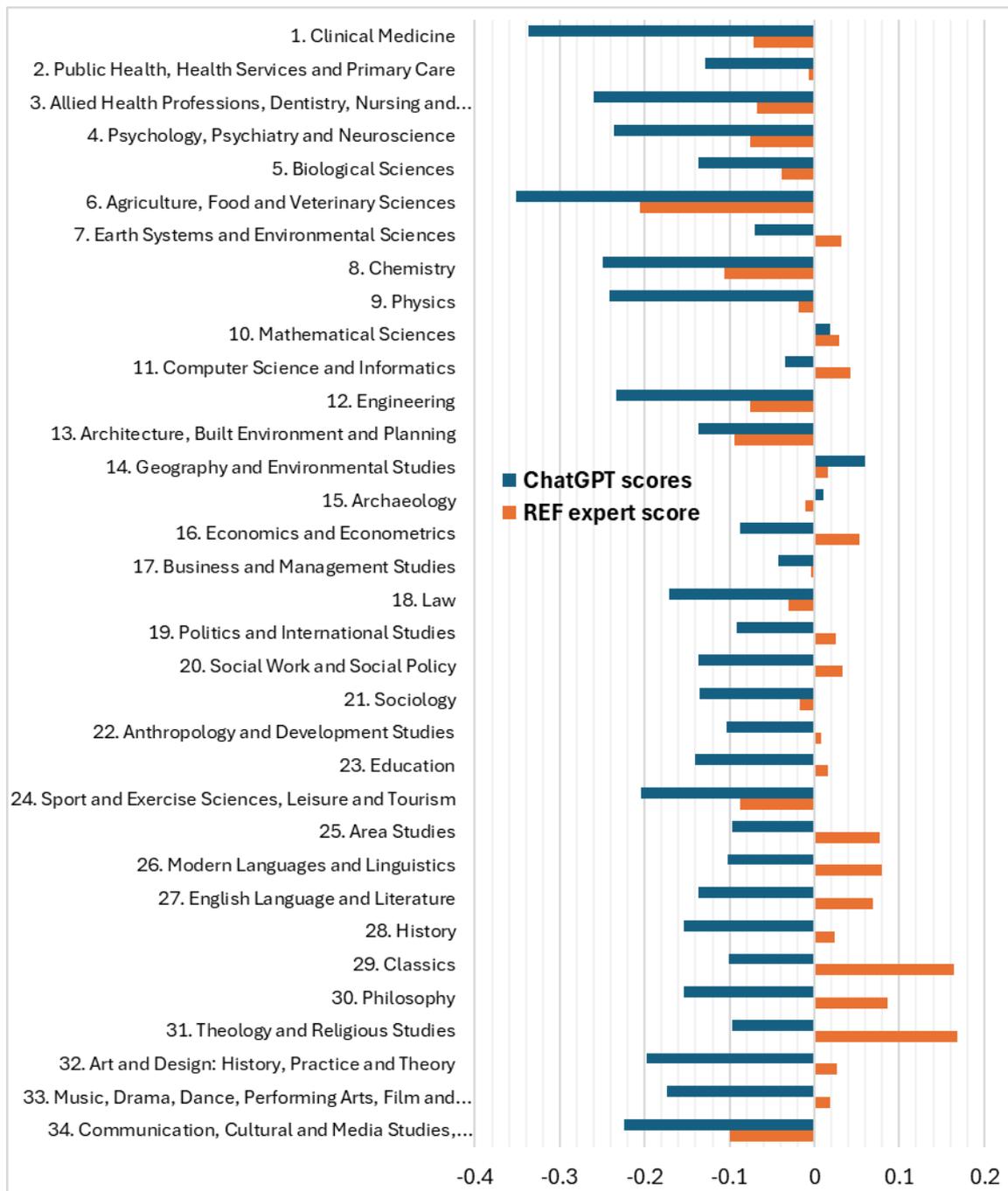

Figure S1. Spearman correlations between **Flesch Reading Ease** with ChatGPT and REF expert scores across the 34 REF Units of Assessment. Higher Flesch Reading Ease scores indicate easier-to-read abstracts (Flesch, 1948). Positive correlations indicate that more readable abstracts tended to receive higher scores, whereas negative correlations indicate that less readable (more linguistically complex) abstracts tended to receive higher scores.

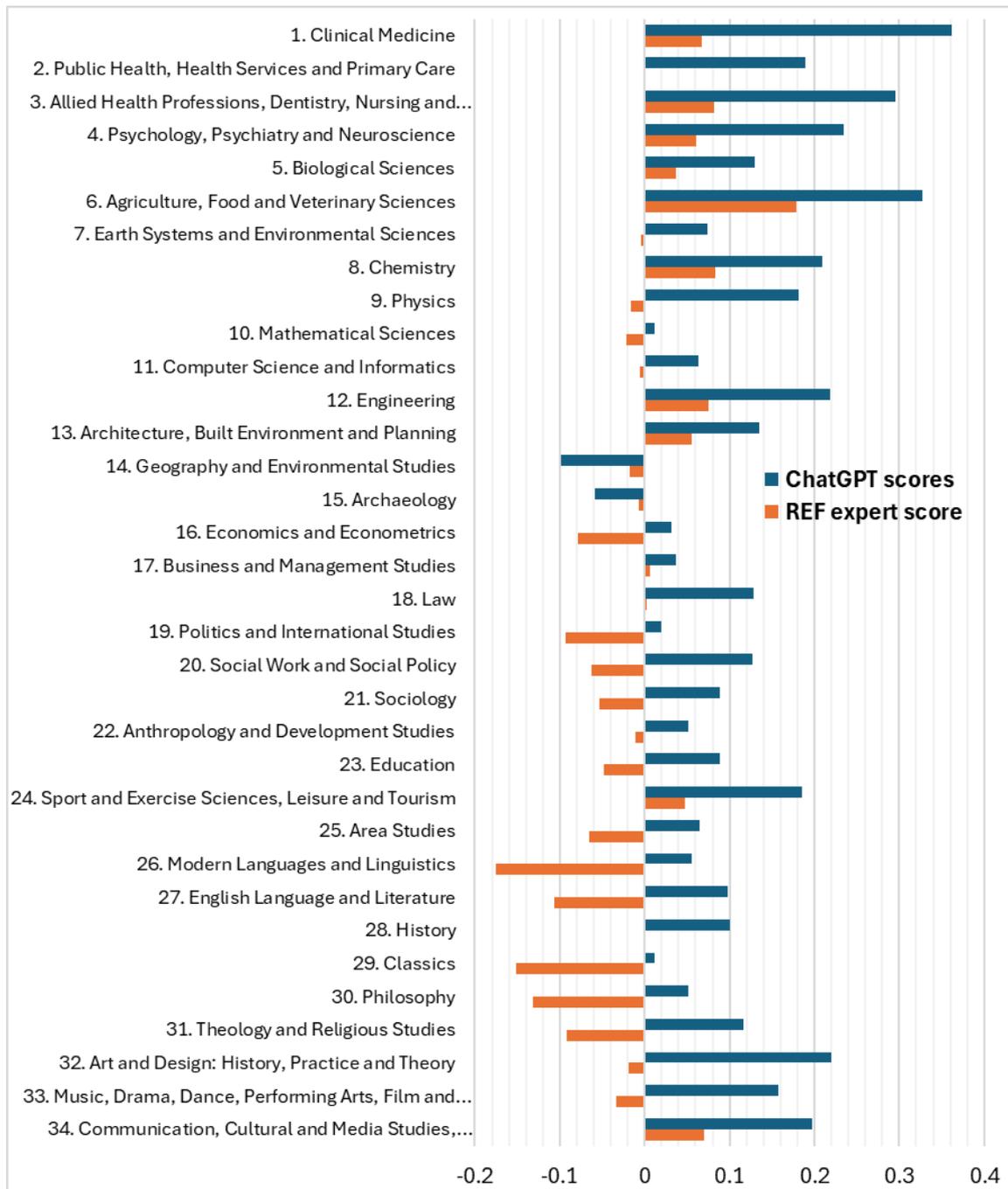

Figure S2. Spearman correlations between **Gunning Fog** readability scores and ChatGPT and REF expert scores across the 34 REF Units of Assessment. Higher Gunning Fog scores indicate more linguistically complex abstracts. Positive correlations indicate that more complex abstracts tended to receive higher scores, whereas negative correlations indicate that simpler abstracts tended to receive higher scores.